\documentclass{article}
\usepackage{spconf,amsmath,graphicx,hyperref}
\usepackage{amsmath, amssymb}
\usepackage{enumitem}
\usepackage{booktabs}
\usepackage{xcolor}
\usepackage{subfig}
\usepackage{enumitem}

\newcommand{\best}[1]{\textbf{#1}}
\newcommand{\secondbest}[1]{\underline{#1}}
\title{HFMCA: Orthonormal Feature Learning for EEG-based Brain Decoding}

\name{Yinghao WANG$^1$\thanks{Thanks to CSC for funding Yinghao WANG and Lintao XU.}, Lintao XU$^2$, Shujian YU$^3$, Enzo TARTAGLIONE$^1$, Van-Tam NGUYEN$^1$}

\address{$^1$LTCI, Télécom Paris, Institute Polytechnique de Paris, Palaiseau, France \\
$^2$LIGM, Univ Gustave Eiffel, École des Ponts, CNRS, Marne-la-Vallée, France \\  
$^3$Department of Artificial Intelligence, Vrije Universiteit Amsterdam, Amsterdam, The Netherlands}

\begin{document}
%\ninept
%
\maketitle
\begin{abstract}
Electroencephalography (EEG) analysis is critical for brain-computer interfaces and neuroscience, but the intrinsic noise and high dimensionality of EEG signals hinder effective feature learning. We propose a self-supervised framework based on the Hierarchical Functional Maximal Correlation Algorithm (HFMCA), which learns orthonormal EEG representations by enforcing feature decorrelation and reducing redundancy. This design enables robust capture of essential brain dynamics for various EEG recognition tasks. We validate HFMCA on two benchmark datasets, SEED and BCIC-2A, where pretraining with HFMCA consistently outperforms competitive self-supervised baselines, achieving notable gains in classification accuracy. Across diverse EEG tasks, our method demonstrates superior cross-subject generalization under leave-one-subject-out validation, advancing state-of-the-art by 2.71\% on SEED emotion recognition and 2.57\% on BCIC-2A motor imagery classification. 
% Through comprehensive ablation studies and analysis, we validate the effectiveness of orthonormal feature learning for EEG recognition task, establishing HFMCA as a powerful framework for decoding brain signals. 
Our code and supplementary material are available at: \url{https://github.com/W-Yinghao/HFMCA_EEG}.
\end{abstract}
\begin{keywords}
Contrastive self-supervised learning, EEG emotion recognition, EEG motor imagery
\end{keywords}
\vspace{-7pt}
\section{Introduction}
\label{sec:intro}

Electroencephalography (EEG) offers a non-invasive window into brain activity, capturing electrical signals that reflect complex neural dynamics across multiple spatiotemporal scales. This modality is indispensable for clinical monitoring and brain-computer interfaces (BCIs), particularly in applications like emotion recognition and motor imagery tasks where physiological signals provide objective biomarkers. However, developing robust and generalizable EEG analysis systems faces significant barriers, primarily stemming from three core challenges. First, the high-dimensional spatiotemporal patterns inherent in multi-channel recordings present difficulties for direct modeling and feature extraction. Second, the analysis is complicated by the extremely low signal-to-noise ratio (SNR), which is further sharpened by substantial inter-subject variability. Third, and critically, the acquisition of large-scale, expertly annotated EEG datasets is prohibitively expensive and time-consuming, creating a severe bottleneck that limits the application of data-hungry fully supervised deep learning approaches.

% ~\cite{wang2025unified}

% necessitating highly noise-resistant methodologies

Contrastive self-supervised learning (SSL) has emerged as a powerful paradigm for leveraging unlabeled EEG data to overcome annotation scarcity. By learning to maximize agreement between differently augmented views of the same instance while minimizing similarity between distinct instances, contrastive SSL encourages the model to extract invariant and semantically meaningful features. This process inherently improves generalization by reducing reliance on spurious correlations~\cite{zhang2022correct} and forcing the encoder to focus on robust, biologically relevant patterns. The resulting representations demonstrate improved transferability across tasks and subjects, and enhanced robustness to inter-session and inter-subject variability, which are key challenges in EEG decoding. Furthermore, by leveraging vast amounts of unlabeled data, contrastive SSL effectively regularizes the model, leading to flatter loss landscapes and better out-of-sample performance~\cite{fradkin2022robustness}. Building on well-known frameworks like SimCLR~\cite{SimCLR} and MoCo~\cite{MoCo}, EEG-based Contrastive SSL methods typically generate positive pairs by applying diverse augmentations to the same raw EEG sample, while treating segments from different samples/trials as negatives. Recent approaches have made progress in addressing domain-specific challenges. Temporal modeling methods such as TS-TCC~\cite{TS-TCC} learn robust representations via cross-view prediction, while spatial modeling approaches like CL-SSTER~\cite{CL_SSTER} use attention to capture inter-channel relationships. For cross-domain generalization, CLISA~\cite{CLISA} addresses domain adaptation in EEG, and multi-modal frameworks~\cite{fan2024leveraging} exploit complementary information across modalities to enhance learning.

%Methods focusing on temporal dynamics, such as TS-TCC~\cite{TS-TCC}, learn robust temporal representations through cross-view prediction mechanisms. Spatial modeling approaches like CL-SSTER~\cite{CL_SSTER} incorporate attention mechanisms to capture inter-channel relationships and spatial patterns. For cross-domain generalization, CLISA~\cite{CLISA} explicitly targets domain adaptation challenges in EEG analysis. Additionally, multi-modal frameworks~\cite{fan2024leveraging} leverage complementary information from multiple data modalities to enhance representation learning. 

% enhance performance in low-data regimes,

However, persistent limitations remain: 1) current frameworks typically decouple temporal and spatial modeling, treating them as independent dimensions rather than jointly learning their complementary representations; 2) conventional negative sampling, which typically relies on random selection across different trials or subjects, is highly prone to semantic contamination given EEG’s low SNR and inter-subject variability. This may group similar brain states as negatives or fail to separate dissimilar, noise-corrupted states~\cite{kan2023self}, thereby degrading representation quality and cross-subject generalizability.

%conventional negative sampling strategies—relying on random selection from different trials or subjects—are acutely vulnerable to semantic contamination due to EEG's low SNR and high inter-subject variability. This can inadvertently group similar brain states as negatives or fail to separate dissimilar states corrupted by noise~\cite{kan2023self}, critically undermining representation quality and cross-subject generalizability, the paramount requirement for real-world BCIs.

%To address these gaps, we propose the Hierarchical Functional Maximal Correlation Algorithm (HFMCA) for self-supervised EEG representation learning. Our framework introduces three core innovations:

This paper presents the Hierarchical Functional Maximal Correlation Algorithm (HFMCA) for self-supervised EEG representation learning and makes three key contributions:
\begin{itemize}[noitemsep,topsep=0pt,parsep=0pt,partopsep=0pt]
    \item We introduce HFMCA in Sec.~\ref{sec:HFMCA}, a hierarchical contrastive self-supervised framework that employs Functional Maximal Correlation Analysis (FMCA)~\cite{hu2022normalized} to maximize dependence between low-level representations (augmentations from more than two views) and the corresponding high-level fused representation~\cite{hu2024learning}, while enforcing orthogonality on encoder outputs. This design is fundamentally different from frameworks like SimCLR or MoCo, where the central idea is to maximize dependence between two augmented parallel views.
    \item We extend HFMCA to HFMCA++ in Sec:~\ref{sec:HFMCA++}, by incorporating an additional regularization term that minimizes the similarity between the averaged low-level features of an instance and the high-level representations of other instances within a batch, which further enhances the discriminative power of the learned representations.
    \item Under leave-one-subject-out validation paradigm, our method demonstrates state-of-the-art cross-subject generalization on both the SEED emotion recognition benchmark and motor imagery classification task, surpassing previous best methods by 2.71\% and 2.54\% in mean accuracy, respectively, described in Sec.~\ref{sec:exp} . %These results collectively highlight the potential for real-world brain signal decoding.
\end{itemize}

% Critically, our method demonstrates notably strong performance in cross-subject validation, achieving state-of-the-art results with an average accuracy of 60.24\% and peak accuracy of 79.05\% on the SEED emotion recognition benchmark under rigorous leave-one-subject-out testing. These results highlight HFMCA's potential as an effective framework for cross-subject EEG modeling, suggesting its utility for developing more generalizable brain signal decoding systems in real-world applications.

% \section{Literature Review}
% \label{sec:format}
\vspace{-7pt}
\section{Methodology}

\begin{figure*}[htbp]
    \centering
    \subfloat[]{\includegraphics[width=0.75\textwidth]{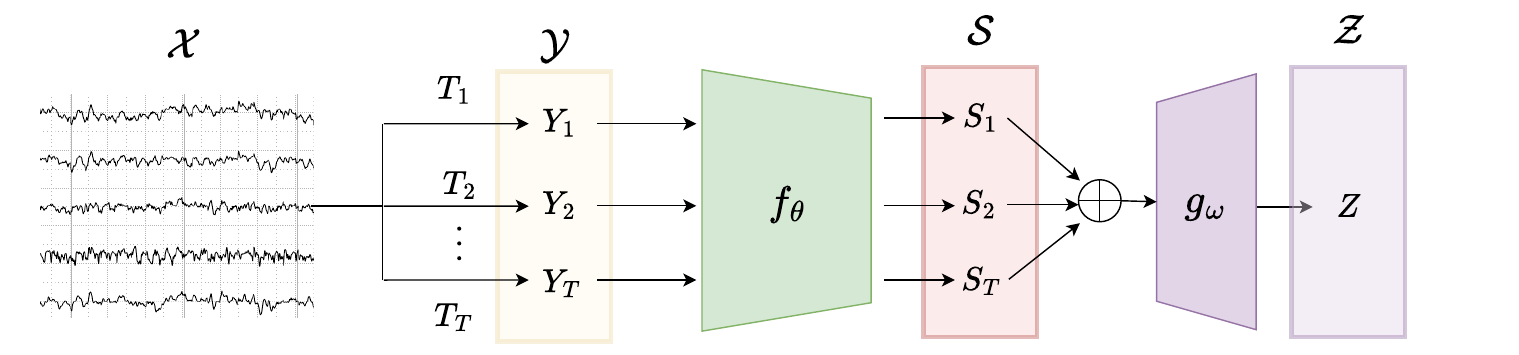}\label{fig:a}}
    % \caption{HFMCA++ architecture}
    \hfill
    \subfloat[]{\includegraphics[width=0.25\textwidth]{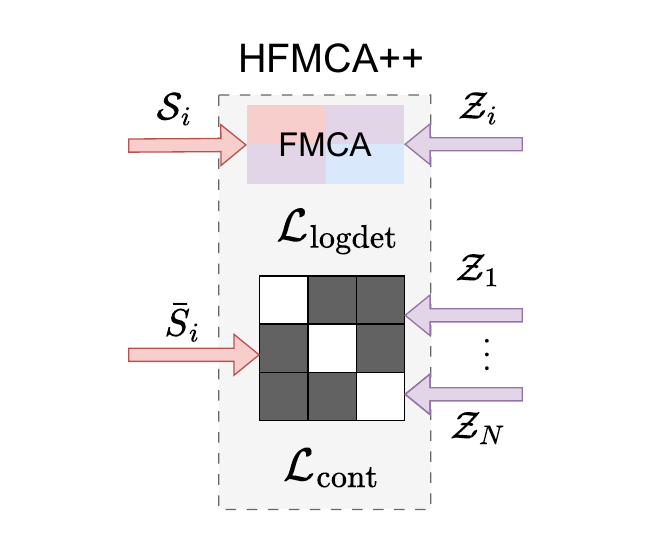}\label{fig:b}}
    % \caption{Hybrid loss structure}
    \caption{(a)~Architecture diagram of the HFMCA framework. The raw EEG signal $\mathcal{X}$ is transformed by functions $T_i$ to generate augmented inputs $\mathcal{Y}$. The encoder $f_\theta$ extracts low-level features $\mathcal{S}$, which are concatenated and fed into $g_\omega$ to produce the high-level instance representation $\mathcal{Z}$.
        (b)~Loss function structure: HFMCA++ employs a hybrid loss $\mathcal{L}_{\text{HFMCA++}}$ composed of the FMCA-based density ratio loss $\mathcal{L}_{\text{logdet}}$ and the cross-instance contrastive loss $\mathcal{L}_{\text{cont}}$.}
    \label{fig:Teaser}
\end{figure*}

% \begin{figure*}
%     \centering
%     \includegraphics[width=0.9\linewidth]{HFMCA++sample.pdf}
%     \caption{caption}
%     \label{fig:Teaser}
% \end{figure*}
\textbf{H}ierarchical \textbf{F}unctional \textbf{M}aximal \textbf{C}orrelation \textbf{A}lgorithm utilizes multiview SSL to investigate the hierarchical relationships between the data and their augmentations.

\subsection{Hierarchical FMCA for Self-Supervised Learning}
\label{sec:HFMCA}
 Similar to other SSL methods, HFMCA utilizes data augmentations to learn meaningful representations. %However, instead of a single pair $(X,Y)$ 
 Let us consider an instance $X$ and a set of augmentations, as described by Eq.~\ref{eq:augumentation}, where $Y_i = T_i(X)$ and $T_i$  signifies $i$-th augmentation function (e.g., temporal masking or channel dropout):
\begin{equation}
    \mathcal{Y} = \{ Y_1, Y_2, \ldots, Y_T \},
\label{eq:augumentation}
\end{equation}
The core concept of our approach posits that each data augmentation generates a distinct low-level representation, capturing a unique perspective of the underlying instance. Aggregating all augmentation-derived low-level features yields a unified high-level representation that summarizes the multiple views and semantically captures the original instance. 
% Whereas conventional methods model statistical dependence $\rho$ between two \textit{specific augmented views} ($X$ and $Y$) -- quantified by how much their joint distribution $P(X,Y)$ deviates from the product of marginals $P(X)P(Y)$ -- our Hierarchical Functional Maximal Correlation Analysis (HFMCA) shifts focus to modeling dependence $\hat{\rho}$ between hierarchical representation levels. 
Specifically, we define $\hat{\rho}$ as the statistical dependence between: (1) the complete \textit{set of low-level features} $\mathcal{S}$ (with variability governed by distribution $P(\mathcal{S})$), and (2) their aggregated \textit{high-level summary} $\mathcal{Z}$ (governed by $P(\mathcal{Z})$). 
%This dependence $\hat{\rho}$, formalized in Eq.~\ref{eq:HFMCA}, measures how significantly the joint distribution $P(\mathcal{S}, \mathcal{Z})$ diverges from $P(\mathcal{S})P(\mathcal{Z})$, thereby quantifying how effectively $\mathcal{Z}$ captures the collective information embedded within $\mathcal{S}$.
This dependence $\hat{\rho}$ is defined as the density ratio between $P(\mathcal{S}, \mathcal{Z})$ and $P(\mathcal{S})P(\mathcal{Z})$, and admits an orthogonal decomposition~\cite{hu2022normalized,huang21universal}:
\begin{equation}
    \hat{\rho}(\mathcal{S}, \mathcal{Z}) = \frac{p(\mathcal{S}, \mathcal{Z})}{p(\mathcal{S})p(\mathcal{Z})} = \sum_{k=1}^{\infty} \sqrt{\sigma_k} \phi_k(\mathcal{S}) \psi_k(\mathcal{Z})
    \label{eq:HFMCA}.
\end{equation}

% representing all augmentation perspectives

%where $\mathcal{S}$ represents features extract from each augmentation, while $\mathcal{Z}$ denotes the high-level feature of the instance.

%The right-hand side of Equation~\ref{eq:HFMCA} represents a spectral decomposition of the density ratio function based on Mercer's Theorem and the theory of reproducing kernel Hilbert spaces (RKHS). 

This decomposition arises from the eigenanalysis of a cross-covariance operator between the reproducing kernel Hilbert spaces (RKHSs) of $\mathcal{S}$ and $\mathcal{Z}$, where $\sigma_k$ denotes the $k$-th eigenvalue quantifying the strength of the $k$-th mode of dependence between $\mathcal{S}$ and $\mathcal{Z}$. The corresponding orthogonal eigenfunctions, $\phi_k(\cdot)$ and $\psi_k(\cdot)$, form complete bases in their respective RKHS, satisfying:
\[
\mathbb{E}_{\mathcal{S}} \left[ \phi_i(\mathcal{S})\phi_{j}(\mathcal{S}) \right] = 
\mathbb{E}_{\mathcal{Z}} \left[ \psi_i(\mathcal{Z})\psi_{j}(\mathcal{Z}) \right] = 
\begin{cases}
1, & i = j \\
0, & i \ne j
\end{cases}
\].

%and their product $\phi_k(\mathcal{S}) \psi_k(\mathcal{Z})$ thus captures the $k$-th mode of joint variation between $\mathcal{S}$ and $\mathcal{Z}$.

% \begin{itemize}[noitemsep,topsep=0pt,parsep=0pt,partopsep=0pt]
%     \item $\mathcal{S} = \{S_i\}_{i=1}^T$ - features from each augumentation, concatenated together.
%     \item $\mathcal{Z}$ - the high-level feature.
% \end{itemize}

To extract hierarchical features from the augmented samples, we employ two neural networks $f_\theta$ and $g_\omega$. Given $T$ augmented views $\{Y_1, Y_2, \dots, Y_T\}$ generated from an input instance, we first extract low-level features using a weight-sharing encoder $f_\theta$ (Fig.~\ref{fig:Teaser}). %This approach enables efficient parameter sharing across augmentations while capturing view-specific characteristics. 
The set of low-level features $\mathcal{S}$ and aggregated high-level representation $\mathcal{Z}$ are obtained by:
% \vspace{-7pt}
\begin{equation}
\begin{aligned}
% \vspace{-7pt}
\mathcal{S} &= \{S_i\}_{i=1}^T= \{f_\theta(Y_1), f_\theta(Y_2), \ldots, f_\theta(Y_T)\} \\
\mathcal{Z} &= g_\omega\left(\text{concat}\left(S_1, S_2, \ldots, S_T\right)\right),
\end{aligned}
\label{eq:imp}
\end{equation}
where $\text{concat}(\cdot)$ denotes the concatenation operation. This hierarchical feature extraction strategy allows $\mathcal{Z}$ to integrate information across all augmentations, forming a comprehensive representation that preserves both view-specific details and global semantics.

Using the low-level feature set $\mathcal{S} = \{S_1, \dots, S_T\}$ and high-level representation $\mathcal{Z}$ from Eq.~\ref{eq:imp}, we compute the correlation matrices as Eq.~\ref{eq:mat}:
\vspace{-7pt}
\begin{equation}
\begin{aligned}
R_1 &= \mathbb{E}\left[\frac{1}{T}\sum_{t=1}^{T} S_t (S_t)^{\top}\right] & 
R_2 &= \mathbb{E}\left[ZZ^{\top}\right] \\
P_{1,2} &= \mathbb{E}\left[\frac{1}{T}\sum_{t=1}^{T} S_t Z^{\top}\right] & 
R_{1,2} &= \begin{bmatrix} R_1 & P_{1,2} \\ P_{1,2}^{\top} & R_2 \end{bmatrix}.
\end{aligned}
\label{eq:mat}
\end{equation}

To maximize the statistical dependence between hierarchical features $\mathcal{S}$ and $\mathcal{Z}$ while promoting feature orthogonality, we minimize a loss function based on log-determinants (logdet) of the correlation matrices~\cite{NIPS,hu2022normalized}. Specifically, we:
\begin{itemize}[noitemsep,topsep=0pt,parsep=0pt,partopsep=0pt]
    \item Minimize $\log \det R_{1,2}$ to enhance dependence between $\mathcal{S}$ and $\mathcal{Z}$.
    \item Maximize $\log \det R_1$ and $\log \det R_2$ to encourage orthogonality within each feature set.
\end{itemize}

This yields the optimization objective shown in Eq.~\ref{eq:imp_logdet}:
\begin{equation}
    \min_{\theta, \omega} \mathcal{L}_{\text{logdet}}  = \underbrace{\log \det R_{1,2}}_{\text{enhance dependence}} - \underbrace{\log \det R_1 - \log \det R_2}_{\text{promote orthogonality}}.
    \label{eq:imp_logdet}
\end{equation}
%where $R_{1,2}$ is the joint correlation matrix of $(\mathcal{S}, \mathcal{Z})$, and $R_1$ and $R_2$ are auto-correlation matrices of $\mathcal{S}$ and $\mathcal{Z}$ respectively. 
Minimizing this objective simultaneously strengthens the relationship between hierarchical representations while encouraging disentangled features within each level.

\subsection{HFMCA++: Preventing Trivial Solutions via Contrastive Regularization}
\label{sec:HFMCA++}
While HFMCA effectively models hierarchical relationships, we observe that the high-level representation $\mathcal{Z}$ is susceptible to over-smoothing or dimensional redundancy, which can cause it to collapse to a trivial solution where all instances map to similar embeddings. To address this limitation, we extend HFMCA to HFMCA++ by incorporating an efficient contrastive regularization term. This novel component minimizes similarity between the averaged low-level features of an instance and the high-level representations of other instances within a batch.
Formally, for a batch of $N$ instances, we first compute the averaged low-level feature for the $i$-th instance:
\vspace{-7pt}
\begin{equation}
\vspace{-7pt}
\bar{S} = \frac{1}{T} \sum_{t=1}^{T} {S_{t}}.
\label{eq:sbar}
\end{equation}
The contrastive regularization term is then defined as:
\begin{equation}
\mathcal{L}_{\text{cont}} = \frac{1}{N(N-1)} \sum_{i=1}^{N} \sum_{j \neq i} \exp\left(\frac{\bar{S_i} \cdot Z_j}{\tau}\right),
\label{eq:lcont}
\end{equation}
where $\tau$ is a temperature hyperparameter controlling similarity scaling.
This contrastive mechanism operates as a repulsive force in the embedding space, ensuring that the high-level representation $Z_i$ maintains discriminative information specific to instance $i$ rather than collapsing to a constant solution. The complete HFMCA++ objective becomes:
\begin{equation}
%\vspace{-7pt}
\min_{\theta, \omega} \mathcal{L}_{\text{HFMCA++}} = \mathcal{L}_{\text{logdet}} + \lambda \mathcal{L}_{\text{cont}},
\end{equation}
with $\lambda\geq 0$ balances maximization of the dependence and the additional regularization term.
% The contrastive term provides three key benefits:
% \begin{enumerate}
%     \item Collapse Prevention: Explicitly penalizes similarity between different instances' representations
%     \item Efficiency: Requires only $O({n^2})$ comparisons rather than $O({n^2}T^2)$ needed by conventional contrastive approaches
%     \item Hierarchy Preservation: Maintains the feature hierarchy by contrasting low-vs-high level features rather than introducing symmetric comparisons 
% \end{enumerate}

The instance-level contrastive loss in HFMCA++ inherently avoids trivial solutions through three mechanisms: (1) augmentation-invariant non-uniformity enforcement via cross-hierarchy asymmetric contrasting of averaged low-level features ($\bar{{S_i}}$) against cross-instance high-level representations ($Z_j$), (2) implicit dimensionality preservation from orthogonal encoder constraints from HFMCA, and (3) cross-hierarchy regularization that filters noise while retaining discriminative variance. Crucially, compared to SimCLR's augmentation-dependent approach, our method reduces inductive bias by leveraging natural inter-instance negatives rather than artificial augmentations. This is particularly vital for EEG data where valid transformations lack theoretical grounding. 

%This strategically optimizes $I(\bar{{S_i}},Z_j)$ rather than $I(S_{aug},S'_{aug}$), eliminating task-irrelevant invariance and explaining its superior generalization.}

% The contrastive term confers three principal advantages: it explicitly prevents representation collapse by penalizing similarity between different instances' embeddings; 
% achieves substantial computational efficiency, requiring only $O({n^2})$ comparisons compared to the $O({n^2}T^2)$ complexity of conventional approaches; 
% and inherently preserves feature hierarchy through asymmetric low-vs-high-level feature contrasting, avoiding symmetry-induced information loss. 
% These properties collectively ensure stable, scalable, and structurally coherent representation learning.
% This extension significantly enhances HFMCA's applicability to real-world EEG analysis where discriminative representations are essential for accurate cross-subject generalization, while maintaining computational efficiency crucial for large-scale neurophysiological datasets.
% After pretraining, $f_\theta^2$ and $f_\theta^3$ are discarded and only the backbone $f_\theta^1$is used for fine-tuning.

% \vspace{-17pt}
\section{Experiments}
\label{sec:exp}
% \subsection{Data Preparation and Augmentation}
\textbf{Datasets:} To evaluate the performance of our proposed method, we conducted extensive experiments on two publicly available and widely used EEG benchmark datasets:
% : the BCIC-2A~\cite{BCIC-2A} dataset for motor imagery classification and the SEED~\cite{SEED} dataset for emotion recognition.
\begin{itemize} [noitemsep,topsep=0pt,parsep=0pt,partopsep=0pt]
    \item \textbf{BCIC-2A~\cite{BCIC-2A}}: contains EEG data from 9 subjects performing four different motor imagery (MI) tasks: imagination of movement of the left hand, right hand, both feet, and tongue. The recordings were acquired using 22 EEG channels at a sampling frequency of 250 Hz. Each subject consists of 72 trials per class (288 trials total). Similarly, we segmented each trial into non-overlapping 4-second segments to create the input samples. The downstream task is a four-class classification problem to identify the imagined movement.
    \item \textbf{SEED~\cite{SEED}}: comprises EEG recordings from 15 subjects collected during emotion elicitation experiments using 15 film clips, resulting in 5 trials per emotion category per subject. The data was recorded using 62 electrodes at a sampling rate of 200 Hz. For our experiments, we segmented each trial into non-overlapping 2-second epochs, resulting in multiple samples per trial for self-supervised pre-training. The final classification task was to predict the emotion label (positive, neutral, or negative) of each trial.
    % Each subject participated in 3 sessions, providing 45 trials per subject. 
\end{itemize}
\textbf{Data Augmentation Strategies:} To learn robust representations in a self-supervised manner, we employ a diverse set of data augmentation techniques specifically designed for EEG signals. Given an input EEG segment $X = (Channels,Time)$, we apply the following transformations: (1) Channel Permutation; (2) Temporal Masking; (3) Channel Dropout, and (4) Temporal Crop-and-Resize.
% \begin{enumerate}
%     \item {Channel Permutation}
%     % : This method permutes the order of EEG channels randomly. This disruption of spatial relationships forces the model to learn features that are invariant to the absolute spatial positioning of electrodes, focusing instead on their functional signals.
%     \item {Temporal Masking}
%     % : Random contiguous portions of the time series are masked (set to zero) across all channels. This technique encourages the model to develop a robust understanding of temporal contexts and not to over-rely on any single salient feature.
%     \item {Channel Dropout}
%     % : Entire channels are randomly selected and set to zero. This simulates the common real-world issue of noisy or disconnected electrodes and compels the model to rely on distributed information from the remaining channels.
%     \item {Temporal Crop-and-Resize}
%     % : A segment of the time series is randomly cropped and then resized back to the original length using interpolation. This augmentation combines warping and cropping, introducing variations in temporal scale and perspective.
% \end{enumerate}
This combination of spatial (channel-based) and temporal augmentations creates a rich set of views for each input, which is crucial for effective contrastive learning. A visualization of the different augmentations and implementation details is provided in the supplementary material.\\
% \subsection{Implementation Details}
\textbf{Network Architecture}: The encoder network $f_\theta$ was adopted from the architecture presented in~\cite{NIPS} for SEED, and in~\cite{li2024self} for BCIC-2A, which are specifically designed for effective EEG representation learning. The projection head $g_\omega$ is implemented as a 3-layer Multilayer Perceptron (MLP) with a hidden layer size of 512 and ReLU activation functions. \\
\textbf{Evaluation Protocol}: Following standard SSL practices, upon completion of the pretext training stage, we freeze the parameters of the pretrained encoder $f_\theta$. A logistic regression (LR) classifier is subsequently trained atop these frozen representations to evaluate the learned embeddings' quality on downstream classification tasks. This protocol provides a rigorous and parameter-efficient assessment of the encoder's representational capacity, ensuring fair comparison with state-of-the-art methods.\\
\textbf{Quantitative Analysis}:
Tables~\ref{tab:seed} and~\ref{tab:bcic} present the experimental results on the SEED  dataset and BCIC-2A dataset, respectively. Overall, HFMCA and HFMCA++ demonstrate superior performance among all compared methods, achieving the second-highest (56.43\%) and highest (58.74\%) average accuracy on SEED, respectively.\\
At the individual subject level, HFMCA++ achieves the best or second-best classification accuracy on 11 out of 15 subjects in the SEED dataset, demonstrating robust cross-subject generalization capability. 
% Notably, HFMCA++ attains remarkable accuracy of 71.45\%, 64.89\%, and 79.05\% on subjects S11, S14, and S15, respectively, showing substantial improvements over baseline methods. Even for challenging subjects (e.g., S2 and S4), HFMCA++ maintains competitive performance compared to other approaches.\\
The results on the BCIC IVa dataset further corroborate these findings. HFMCA++ consistently achieves the highest average accuracy and obtains the best or second-best performance across the majority of subjects. This cross-dataset consistency demonstrates that our proposed method is not only effective for emotion recognition but also generalizes well to other EEG tasks such as motor imagery classification. Such task-agnostic performance validates the robustness of the HFMCA framework.\\
Compared to conventional self-supervised learning methods, HFMCA achieves an improvement of approximately 4--8 percentage points, which represents a significant advancement in EEG signal analysis. This substantial performance gain can be attributed to our hierarchical feature modeling, which effectively captures both temporal dynamics and inter-channel relationships in EEG signals.
\renewcommand{\arraystretch}{1.0}
\begin{table}[h]
  \setlength{\tabcolsep}{2.5pt}
  \caption{Comparative results on SEED emotion recognition dataset. We report the classification accuracy (\%) of our proposed HFMCA and  HFMCA++ against five baseline methods, including MoCo~\cite{MoCo}, SimCLR~\cite{SimCLR}, BYOL~\cite{BYOL}, SimSiam~\cite{Simsiam} and ContraWR~\cite{ContraWR}.The best results are shown in bold, while the second-best results are underlined.} 
  \label{tab:seed}
  \centering
  \small
  \scalebox{0.8}{
  \begin{tabular} {l|ccccc|cc} 
      \toprule 
      \textbf{Subject}   & \textbf{MoCo} & \textbf{SimCLR} & \textbf{BYOL} & \textbf{SimSiam} & \textbf{ContraWR}  & \textbf{HFMCA} & \textbf{HFMCA++}  \\
      \midrule
      S1  & 40.69 & 44.17 & 42.41 & 41.63 & 48.56 & \secondbest{50.92} & \best{53.54} \\
      S2  & 40.48 & \best{48.70} & 39.52 & 42.25 & 42.58 & \secondbest{48.25} & 47.88 \\

      S3  & 45.96 & \best{61.99} & 52.19 & 47.84 & 56.45 & 59.38 & \secondbest{59.73} \\
      S4  & \best{51.31} & 49.03 & 50.08 & 45.61 & \secondbest{51.29} & 47.14 & 49.61 \\
      S5  & 47.54 & 50.77 & \secondbest{52.78} & 48.07 & 49.13 & 49.91 & \best{53.15} \\
      S6  & 61.53 & 59.05 & \best{64.71} & 62.21 & 61.98 & 60.49 & \secondbest{62.55} \\
      S7  & 39.27 & \secondbest{49.70} & 44.15 & \best{51.73} & 46.96 & 49.62 & 49.60 \\
      S8  & 60.80 & 61.92 & 60.06 & 54.89 & \secondbest{63.58} & 60.57 & \best{64.07} \\
      S9  & 57.54 & 58.30 & \best{59.36} & 56.45 & 58.25 & 56.18 & \secondbest{59.11} \\
      S10 & 50.40 & 56.62 & 53.25 & \best{59.40} & 55.32 & 56.63 & \secondbest{58.73} \\
      S11  & 58.54 & \secondbest{67.51} & 64.65 & 61.49 & 66.45 & 67.35 & \best{71.45}\\
      S12 & 45.52 & 49.03 & \secondbest{49.66} & \best{51.14} & \secondbest{49.66} & 46.44 & 49.52  \\
      S13 & 54.03 & 57.66 & 60.20 & 56.21 & 58.46 & \secondbest{61.27} & \best{64.89} \\
      S14 & 52.47 & 60.14 & 55.73 & 57.33 & 59.07 & \secondbest{60.30} & \best{63.54} \\
      S15  & 56.62 & 66.11 & 60.06 & 64.59 & 57.85 & \secondbest{72.01} & \best{73.74} \\
      \midrule
      Avg   & 50.84 & 56.03 & 53.92 & 53.39 & 55.04 & \secondbest{56.43} & \best{58.74}  \\
      \bottomrule 
  \end{tabular}
  }
\end{table}
\renewcommand{\arraystretch}{1.0}
\begin{table}[htp]
  \setlength{\tabcolsep}{2.5pt}
  \caption{Comparative results on BCIC-2A motor imagery classification dataset.} 
  \label{tab:bcic}
  \centering
  \small
  \scalebox{0.8}{
  \begin{tabular} {l|ccccc|cc} 
      \toprule 
      \textbf{Subject}  & \textbf{MoCo} & \textbf{SimCLR} & \textbf{BYOL} & \textbf{SimSiam} & \textbf{ContraWR}  & \textbf{HFMCA} & \textbf{HFMCA++}  \\
      \midrule 
      S1   & 39.58 & 39.75 & 35.94 & 35.24 & 40.10 & \best{41.49} & \secondbest{40.97}   \\
      S2   & 39.58 & 41.32 & 38.02 & 41.49 & 43.05 & \secondbest{43.40} & \best{44.10}  \\
      S3   & 42.36 & 43.58 & 41.67 & 47.05 & 47.74 & \best{50.86} & \secondbest{50.69}  \\
      S4   & 37.33 & 37.33 & 35.76 & 36.63 & 38.54 & \best{39.58} & \secondbest{39.06}  \\
      S5   & 39.24 & 47.22 & 43.75 & 43.92 & 47.04 & \best{52.60} & \best{52.60} \\
      S6   & 39.24 & 39.76 & 35.93 & 33.68 & 37.50 & \secondbest{42.36} & \best{43.23} \\
      S7   & 42.53 & 48.78 & 43.40 & 43.23 & 50.35 & \secondbest{55.38} & \best{55.56} \\
      S8   & 39.41 & 40.97 & 38.19 & 40.62 & 41.84 & \secondbest{44.27} & \best{47.56} \\
      S9   & 42.88 & 44.27 & 42.53 & \secondbest{44.79} & \best{45.83} & 40.79 & 41.32 \\
      \midrule
      Avg   & 40.24 & 42.89 & 38.35 & 40.73 & 43.55 & \secondbest{45.63} & \best{46.12}   \\
      \bottomrule 
  \end{tabular}
  }
\end{table}
% \vspace{-17pt}
\section{Conclusion And Future Work}
\label{sec:conclu}
In this work, we presented HFMCA, a novel hierarchical self-supervised learning framework for EEG representation learning that effectively balances feature dependence and discriminative power. By combining log-determinant minimization for hierarchical feature alignment with an efficient contrastive regularization mechanism, our approach prevents representation collapse while maintaining computational efficiency. Validation on SEED and BCIC-2A dataset demonstrates state-of-the-art performance, with HFMCA++ achieving  2.71\% and 2.57\% higher accuracy compared to conventional contrastive methods, respectively. The learned representations show enhanced discriminative power and improved cross-subject generalization capabilities, critical for real-world brain-computer interface applications. Future work will explore: 
\begin{itemize}[noitemsep, topsep=0pt]
    \item Extending the hierarchical framework to incorporate temporal dynamics of EEG signals,
    \item Investigating domain adaptation capabilities for cross-dataset transfer learning,
    \item Exploring HFMCA to develop EEG foundation models by leveraging large-scale data from diverse cohorts.
\end{itemize}

\section{Acknowledgement}
This work was supported by the French National Research Agency (ANR) in the framework of the IA Cluster project “Hi! PARIS Cluster 2030” under Grant ANR-23-IACL-005, and by the Hi! PARIS Center on Data Analytics and Artificial Intelligence.

%\clearpage
%\vfill\pagebreak
% \newpage

\bibliographystyle{IEEEbib}
\bibliography{strings,refs}

@inproceedings{MoCo,
  title={Momentum contrast for unsupervised visual representation learning},
  author={He, Kaiming and Fan, Haoqi and Wu, Yuxin and Xie, Saining and Girshick, Ross},
  booktitle={CVPR},
  pages={9729--9738},
  year={2020}
}

@inproceedings{SimCLR,
  title={A simple framework for contrastive learning of visual representations},
  author={Chen, Ting and Kornblith, Simon and Norouzi, Mohammad and Hinton, Geoffrey},
  booktitle={International conference on machine learning},
  pages={1597--1607},
  year={2020}
}

@article{BYOL,
  title={Bootstrap your own latent-a new approach to self-supervised learning},
  author={Grill, Jean-Bastien and others},
  journal={Advances in neural information processing systems},
  volume={33},
  pages={21271--21284},
  year={2020}
}

@inproceedings{Simsiam,
  title={Exploring simple siamese representation learning},
  author={Chen, Xinlei and He, Kaiming},
  booktitle={Proceedings of the IEEE/CVF conference on computer vision and pattern recognition},
  pages={15750--15758},
  year={2021}
}

@article{ContraWR,
  title={Self-supervised electroencephalogram representation learning for automatic sleep staging: model development and evaluation study},
  author={Yang, Chaoqi and Xiao, Cao and Westover, M Brandon and Sun, Jimeng},
  journal={JMIR AI},
  volume={2},
  number={1},
  pages={e46769},
  year={2023}
}

@article{kan2023self,
  title={Self-supervised group meiosis contrastive learning for EEG-based emotion recognition},
  author={Kan, Haoning and Yu, Jiale and Huang, Jiajin and Liu, Zihe and Wang, Heqian and Zhou, Haiyan},
  journal={Applied Intelligence},
  volume={53},
  number={22},
  pages={27207--27225},
  year={2023}
}

@article{li2024self,
  title={Self-supervised contrastive learning for EEG-based cross-subject motor imagery recognition},
  author={Li, Wenjie and Li, Haoyu and Sun, Xinlin and Kang, Huicong and An, Shan and Wang, Guoxin and Gao, Zhongke},
  journal={Journal of Neural Engineering},
  volume={21},
  number={2},
  pages={026038},
  year={2024},
  publisher={IOP Publishing}
}

@article{TS-TCC,
  title={Self-supervised contrastive representation learning for semi-supervised time-series classification},
  author={Eldele, Emadeldeen and others},
  journal={IEEE Transactions on Pattern Analysis and Machine Intelligence},
  volume={45},
  number={12},
  pages={15604--15618},
  year={2023},
  publisher={IEEE}
}

@article{CLISA,
  title={Contrastive learning of subject-invariant EEG representations for cross-subject emotion recognition},
  author={Shen, Xinke and Liu, Xianggen and Hu, Xin and Zhang, Dan and Song, Sen},
  journal={IEEE Transactions on Affective Computing},
  volume={14},
  number={3},
  pages={2496--2511},
  year={2022},
  publisher={IEEE}
}

@inproceedings{fan2024leveraging,
  title={Leveraging contrastive learning and self-training for multimodal emotion recognition with limited labeled samples},
  author={Fan, Qi and Li, Yutong and Xin, Yi and Cheng, Xinyu and Gao, Guanglai and Ma, Miao},
  booktitle={Proceedings of the 2nd International Workshop on Multimodal and Responsible Affective Computing},
  pages={72--77},
  year={2024}
}

@article{CL_SSTER,
  title={Contrastive learning of shared spatiotemporal EEG representations across individuals for naturalistic neuroscience},
  author={Shen, Xinke and Tao, Lingyi and Chen, Xuyang and Song, Sen and Liu, Quanying and Zhang, Dan},
  journal={NeuroImage},
  volume={301},
  pages={120890},
  year={2024},
  publisher={Elsevier}
}

@article{BCIC-2A,
  title={Review of the BCI competition IV},
  author={Tangermann, Michael and others},
  journal={Frontiers in neuroscience},
  volume={6},
  pages={55},
  year={2012},
  publisher={Frontiers Research Foundation}
}

@article{SEED,
  title={Investigating critical frequency bands and channels for EEG-based emotion recognition with deep neural networks},
  author={Zheng, Wei-Long and Lu, Bao-Liang},
  journal={IEEE Transactions on autonomous mental development},
  volume={7},
  number={3},
  pages={162--175},
  year={2015},
  publisher={IEEE}
}

@article{NIPS,
  title={Learning cortico-muscular dependence through orthonormal decomposition of density ratios},
  author={Ma, Shihan and Hu, Bo and Jia, Tianyu and Clarke, Alexander and Zicher, Blanka and Caillet, Arnault and Farina, Dario and Pr{\'\i}ncipe, Jos{\'e} C},
  journal={NeurIPS},
  volume={37},
  pages={129303--129328},
  year={2024}
}

@inproceedings{zhang2022correct,
  title={Correct-N-Contrast: a Contrastive Approach for Improving Robustness to Spurious Correlations},
  author={Zhang, Michael and Sohoni, Nimit S and Zhang, Hongyang R and Finn, Chelsea and Re, Christopher},
  booktitle={ICML},
  pages={26484--26516},
  year={2022}
}

@inproceedings{fradkin2022robustness,
  title={Robustness to adversarial gradients: A glimpse into the loss landscape of contrastive pre-training},
  author={Fradkin, Philip and Atanackovic, Lazar and Zhang, Michael R},
  booktitle={ICML Workshop},
  year={2022}
}

@article{huang21universal,
  title={Universal Features for High-Dimensional Learning and Inference},
  author={Huang, Shao-Lun and Makur, Anuran and Wornell, Gregory W and Zheng, Lizhong},
  journal={Foundations and Trends{\textregistered} in Communications and Information Theory},
  volume={21},
  number={1-2},
  pages={1--299},
  year={2024}
}

@inproceedings{hu2024learning,
  title={Learning Orthonormal Features in Self-Supervised Learning using Functional Maximal Correlation},
  author={Hu, Bo and Bu, Yuheng and Pr{\'\i}ncipe, Jos{\'e} C},
  booktitle={IEEE International Conference on Image Processing},
  pages={472--478},
  year={2024}
}

@article{hu2022normalized,
  title={The normalized cross density functional: A framework to quantify statistical dependence for random processes},
  author={Hu, Bo and Principe, Jose C},
  journal={arXiv preprint arXiv:2212.04631},
  year={2022}
}

\end{document}